\documentclass[manuscript]{acmart}
\AtBeginDocument{%
  \providecommand\BibTeX{{%
    \normalfont B\kern-0.5em{\scshape i\kern-0.25em b}\kern-0.8em\TeX}}}

\copyrightyear{2024}
\acmYear{2024}
\setcopyright{rightsretained}
\acmConference[CHIIR '24]{Proceedings of the 2024 ACM SIGIR Conference on Human Information Interaction and Retrieval}{March 10--14, 2024}{Sheffield, United Kingdom}
\acmBooktitle{Proceedings of the 2024 ACM SIGIR Conference on Human Information Interaction and Retrieval (CHIIR '24), March 10--14, 2024, Sheffield, United Kingdom}\acmDOI{10.1145/3627508.3638344}
\acmISBN{979-8-4007-0434-5/24/03}

\usepackage{multirow}
\usepackage{array}
\begin{document}

%%
%% The "title" command has an optional parameter,
%% allowing the author to define a "short title" to be used in page headers.
\title{Task Supportive and Personalized Human-Large Language Model Interaction: A User Study}

\author{Ben Wang}
\email{benw@ou.edu}
\orcid{0000-0001-8612-1185}
\author{Jiqun Liu}
\email{jiqunliu@ou.edu}
\author{Jamshed	Karimnazarov}
\email{jamshed.k@ou.edu}
\author{Nicolas Thompson}
\email{nicolas.f.thompson-1@ou.edu}
\affiliation{%
  \institution{The University of Oklahoma}
  \city{Norman}
  \state{Oklahoma}
  \country{USA}
}

\renewcommand{\shortauthors}{Wang, et al.}

%%
%% The abstract is a short summary of the work to be presented in the
%% article.
\begin{abstract}
Large language model (LLM) applications, such as ChatGPT, are a powerful tool for online information-seeking (IS) and problem-solving tasks. However, users still face challenges initializing and refining prompts, and their cognitive barriers and biased perceptions further impede task completion. These issues reflect broader challenges identified within the fields of IS and interactive information retrieval (IIR). To address these, our approach integrates \textit{task context} and \textit{user perceptions} into human-ChatGPT interactions through prompt engineering. We developed a ChatGPT-like platform integrated with supportive functions, including perception articulation, prompt suggestion, and conversation explanation. Our findings of a user study demonstrate that the supportive functions help users manage expectations, reduce cognitive loads, better refine prompts, and increase user engagement. This research enhances our comprehension of designing proactive and user-centric systems with LLMs. It offers insights into evaluating human-LLM interactions and emphasizes potential challenges for under served users.
\end{abstract}



%%
%% Keywords. The author(s) should pick words that accurately describe
%% the work being presented. Separate the keywords with commas.

%%
%% This command processes the author and affiliation and title
%% information and builds the first part of the formatted document.
\maketitle
© Wang et al., 2024. This is the author's version of the work. It is posted here for your personal use. Not for redistribution. The definitive Version of Record was published in ACM CHIIR 2024, https://doi.org/10.1145/3627508.3638344.

\section{Introduction}
The release of ChatGPT has sparked considerable interest in the interaction between humans and AI. This interest has led to a rising number of individuals employing large language models (LLMs) for various purposes such as task assistance, entertainment, education, and even as an alternative to traditional search engines \cite{bahrini2023chatgpt, brown2020language}. Despite the prevalence of ChatGPT, users still face challenges formulating prompts, and cognitive barriers and biased perceptions further impede task completion \cite{skjuve2023user, zamfirescu2023johnny}. These issues reflect broader challenges identified within the fields of information seeking (IS) and interactive information retrieval (IIR), particularly concerning task context and user perceptions, such as task topic and type, user intent, topic familiarity, and task expectations \cite{azzopardi2021cognitive, liu2020identifying, liu2022leveraging, savolainen2015cognitive, wang2022investigating}. 

Previous IIR studies have underscored the complexity of integrating these fluctuating user perceptions into a predominantly static search system. Fortunately, the evolution of LLMs marks a transformative phase in information access (IA) paradigms, introducing a promising avenue for incorporating more nuanced interaction data through conversational context between the user and generative IA (GIA) system \cite{shah2023envisioning}. 
% When evaluating the human-LLM interaction concerning user experience and task completion, particularly in prompt formulation, user engagement, and learning outcome, 
Therefore, it becomes crucial to explore methodologies for embedding task context and user perceptions into ChatGPT interactions, subsequently evaluating their impact on user experience and task completion, which are essential criteria for evaluation IIR and Human-AI Interaction \cite{liu2018satisfaction, odijk2015struggling, amershi2019guidelines, skjuve2023user}. 
% In light of this, our research pursues two primary objectives: \textbf{identifying the specific tasks users aim to perform with ChatGPT, and determining the forms of support that can effectively mitigate their challenges}.

To achieve this, we have developed a task platform that emulates the official ChatGPT interface, incorporating the GPT-3.5-turbo model. Aiming to support users with the challenges mentioned above, we designed and implemented three supportive functions: 1. \textbf{Perception Articulation}: allows users to clarify their perceptions, including topic familiarity, and expected task complexity. This perception articulation will be then input to ChatGPT through prompt engineering to enrich the context information; 2. \textbf{Prompt Suggestions}: generates prompt revisions and follow-up questions, aiding users who struggle with prompt formulation; 3. \textbf{Conversation Explanation}: generates explanations for the ongoing conversation (i.e., the user's prompt and ChatGPT's response pair) for users to better comprehend ChatGPT's interpretation of the conversation.

To validate our approach, we conducted a naturalistic user study, involving 16 participant of college students and crowdsourced workers with self-defined tasks. These tasks spanned various lengths and cover diverse topics including creative writing, professional development, and specific programming questions.

Our analysis underscores the effectiveness of the supportive functions, illuminating their role in facilitating user experience and task completion. The findings reveal that these functions proved instrumental in managing user expectations, reducing cognitive load, guiding prompt refinement, and increasing user engagement. This research further enhances our understanding of designing proactive and user-centric systems with LLMs, offering insights into evaluating human-LLM interactions from both the system and user ends, and underscoring potential challenges for under served users in this new era of AI.

\section{Related work}
\subsection{Capability of ChatGPT}
LLMs have emerged as a groundbreaking development in the realm of artificial intelligence, leveraging sophisticated architectures trained on extensive data to understand and emulate human-like text generation \cite{ouyang2022training}. One such application is ChatGPT, which has seen the most significant growth in its userbase. One key access to the versatility of LLMs is prompt engineering, a method that uses specific information and instruction in the input to optimize ChatGPT's output content and format \cite{liu2023pre}. Previous studies have examined using ChatGPT and other LLMs in educational settings where they can personalize content delivery and foster enhanced learning experiences \cite{baidoo2023education, kasneci2023chatgpt, tlili2023if, zhai2022chatgpt}. In addition, ChatGPT has demonstrated potential in providing emotional support, by playing therapeutic roles based on user sentiment and need \cite{tlili2023if}. 

\subsection{LLMs as Information Access Systems}
Incorporating LLMs in information access systems introduces transformative prospects from GIA, particularly through multi-turn interactions that resemble traditional IIR processes \cite{shah2023envisioning, skjuve2023user}. However, harnessing LLMs in information access systems presents pronounced challenges. Users encounter difficulties in query formulation or interpreting search results, often due to cognitive barriers, which are common when initializing and refining prompts during interactions with LLMs \cite{zamfirescu2023johnny, savolainen2015cognitive}. These barriers often stem from task context and user perceptions, such as lack of prior knowledge, low familiarity level with the topic, high complexity, and inappropriate expectations, leading to potential misconceptions about how LLMs interpret and respond \cite{odijk2015struggling, hassan2014struggling, wang2022investigating, wang2023investigating, savolainen2015cognitive, ford1995information}.

\subsection{Evaluating Human-LLM-Interaction}
In recent literature, the evaluation of human-LLM interaction has garnered significant attention, especially as these models become increasingly used in human tasks. A comprehensive approach to this evaluation has been proposed, encompassing aspects such as task performance, user experience, and general "Human-AI eXperience" (HAX) \cite{amershi2019guidelines,lee2022evaluating}. With similar interaction process and challenges, there are also notable parallels between the evaluation of human-LLM interaction and the assessment of IIR. Both areas highlight the significance of user experience and perception. Another critical facet enhancing user trust and comprehension is explainability in AI, which merits more profound exploration \cite{xu2019explainable, kim2023help}.

\section{Research Questions}
In respect to the challenges and opportunities in human-LLM interaction and the roles of task context and user perceptions in IS and IIR research, our study aims to investigate the research question \textbf{RQ}: How can we provide support with task context and user perception information to mitigate user challenges in tasks when interacting with ChatGPT?

To answer this question, we explore our methodology for collecting and integrating features related to task context and user perception, importing these features into the system through prompt engineering, and developing supportive functions to enhance user assistance in the task.

\section{Methodology}
\subsection{System design and supportive functions}
\begin{figure*}
\centering
\centering
\includegraphics[width=0.8\linewidth]
{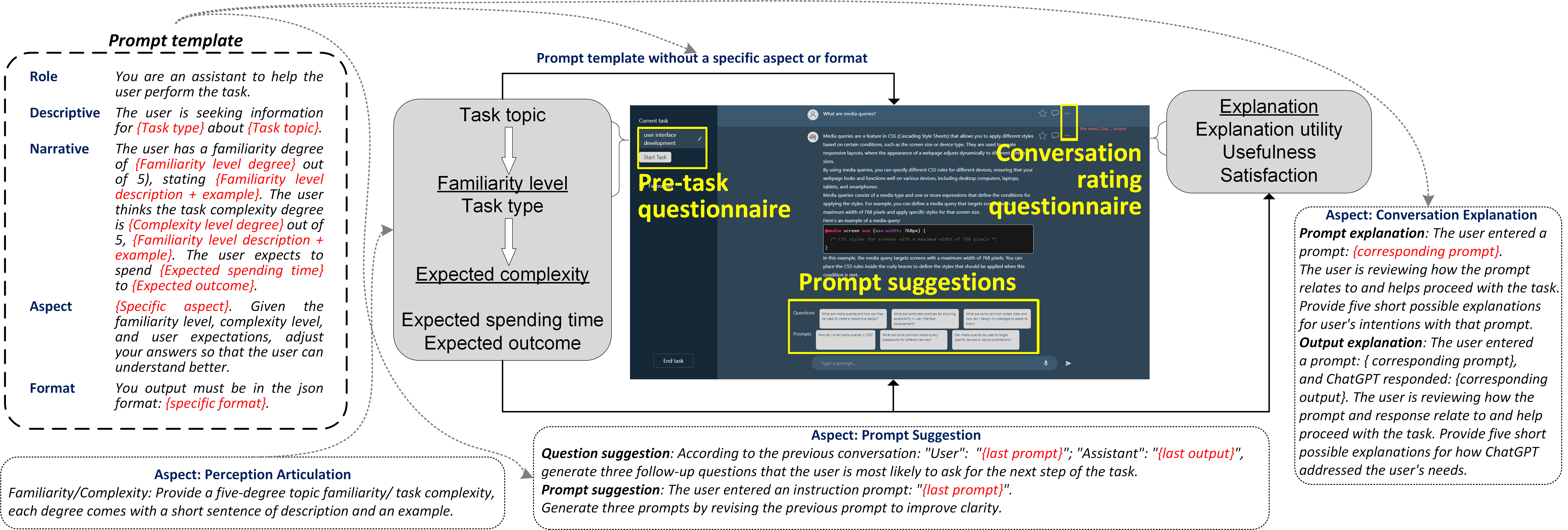}  
\caption{User study platform and prompt templates for supportive functions. Yellow boxes highlight the components for the questionnaires and supportive functions. Grey boxes contain features (including \underline{generative features}) collected through the questionnaires. Solid arrows indicate the features collected in the pre-task questionnaire, subsequently utilized in prompt suggestions and conversation explanations through prompt engineering. Dashed and dotted boxes contain prompt templates, with \textcolor{red}{\textit{\{variable features\}}}. Dotted arrows indicate the application of prompt templates in implementing the supportive functions.}
\label{fig:process}
\end{figure*}

We developed an interface similar to the official ChatGPT, with GPT-3.5-turbo as the model, and integrated questionnaires and supportive functions. The interface and the prompt templates for the supprortive functions are shown in Figure \ref{fig:process}. To interact with the system, users need to enter the pre-task questionnaire highlighted in upper left of the interface. The pre-task questionnaire collects features of task context and user perceptions, including task topic and type, familiarity level, and expectations (e.g., expected task complexity, spending time, and outcome). The detailed descriptions of these features are in Table \ref{tab:fea}. Within the pre-task questionnaire, perceptions articulation is implemented as the generative features: familiarity level and expected complexity. Unlike traditional surveys that use Likert scales to measure these two features, our approach utilizes ChatGPT to generate five degrees of familiarity or task complexity with descriptions and examples. This function aims to assist participants in better comprehending and selecting the option that most closely aligns with their perceptions. The chosen degree, along with its description and example, are then formatted in the prompt template to enrich the context of the main prompt template, which is demonstrated in the left dashed box in Figure \ref{fig:process}. We developed this main prompt template with several components, including role, description, narrative, aspect, and format according to a prior study for designing effective prompts \cite{thomas2023large}. This template aims to provide ChatGPT with comprehensive background information about task context and user perceptions.

After this pre-task questionnaire, we implement the second supportive function, prompt suggestions, by using the main prompt template involved ongoing conversations. The suggestions are displayed in separate tabs at the bottom of the interface. Furthermore, for each conversation, we implement the conversation explanation function in the rating questionnaire. This function generates five explanation options, allowing users to select the one that best aligns with their intent. We also include an explanation utility question, which allows users to rate the utility of the chosen explanation on a five-point scale. The purpose of the conversation explanation function is to present ChatGPT's interpretation of each prompt or response for users and investigate the potential of these explanations in enhancing user engagement and experience.

\begin{table}
\caption{Features in the pre-task questionnaire and the conversation rating questionnaire.}
\label{tab:fea}
\centering
\footnotesize
\renewcommand{\arraystretch}{0.9}
\begin{tabular}{l|l} \hline
Feature & Description \\ \hline
\multicolumn{2}{l}{\textit{Pre-task questionnaire ~}} \\ \hline
Task topic & User input text of task topic \\ \hline
Task type & \begin{tabular}[c]{@{}l@{}}General task type options, such as Learning a \\new topic, generating text, casual conversation, \\replacing search engine, etc.\end{tabular} \\ \hline
\begin{tabular}[c]{@{}l@{}}Familiarity \\level\end{tabular} & \begin{tabular}[c]{@{}l@{}}Five-degree choices with descriptions and examples \\generated by a prompt involving \{task topic\}\end{tabular} \\ \hline
\begin{tabular}[c]{@{}l@{}}Expected \\complexity\end{tabular} & \begin{tabular}[c]{@{}l@{}}Five-degree choices with descriptions and examples \\generated by a prompt involving \{Task topic\} \\and \{Familiarity level\}\end{tabular} \\ \hline
\begin{tabular}[c]{@{}l@{}}Expected \\spending time\end{tabular} & \begin{tabular}[c]{@{}l@{}}Three options: less than 30 minutes, 1 to 2 hours, \\more than 3 hours\end{tabular} \\ \hline
\begin{tabular}[c]{@{}l@{}}Expected \\outcome\end{tabular} & \begin{tabular}[c]{@{}l@{}}Four options: get a broad idea of the task, \\partially complete the task, \\fully complete the task, other(text input)\end{tabular} \\ \hline
\multicolumn{2}{l}{\textit{Conversation rating questionnaire}} \\ \hline
Explanation & \begin{tabular}[c]{@{}l@{}}Five potential explanations for corresponding \\\{user prompt\} and \{ChatGPT response\}\end{tabular} \\ \hline
\begin{tabular}[c]{@{}l@{}}Explanation \\utility\end{tabular} & \begin{tabular}[c]{@{}l@{}}Five-degree options with descriptions from \\“very poor” to “excellent”\end{tabular} \\ \hline
\begin{tabular}[c]{@{}l@{}}Conversation \\usefulness\end{tabular} & \begin{tabular}[c]{@{}l@{}}Five-degree options from not useful to \\extremely useful.\end{tabular} \\ \hline
\begin{tabular}[c]{@{}l@{}}Conversation \\satisfaction\end{tabular} & \begin{tabular}[c]{@{}l@{}}Five-degree options from very unsatisfied to \\very satisfied.\end{tabular} \\ \hline
\end{tabular}
\end{table}

\subsection{Participant recruitment}
We targeted two distinct user groups for our research: college students at a research university and crowdsourced workers from Amazon mTurk. The recruitment process contains two steps: \textbf{Step 1}: participants were asked to complete a registration survey. This survey collected background information, including demographics and prior experience with ChatGPT. \textbf{Step 2}: We then inquired participants if they wished to proceed to the remote user study. Those who opted in then reported the tasks they planned to perform with ChatGPT. We required participants to report task plans with three anticipated task lengths: short (less than 30 minutes), medium (1 to 2 hours), and long (3 hours or more). According to the naturalistic study setting, participants were allowed to edit the task plan when they had new task ideas and complete planned tasks in five days. Before their own tasks, they would perform a warm-up task to get familiar with the platform interface and the study process. After the user study, participants could opt for an interview where we sought their feedback and insights on their experiences. In these interviews, we specifically explored their views on the task experience using our platform, the effectiveness of the supportive functions, and any suggestions or opinions they might have. Compensation for participants includes \$5 for the step 1 registration survey and \$50 for the step 2 user study. This compensation exceeds the minimum wage threshold, and our research has received approval from the Institutional Review Board (IRB).

The decision to choose two distinct participant groups aimed at broadening user diversity. While past studies focused on college students as early adopters of ChatGPT, they still highlighted the need for a more heterogeneous user group. Consequently, our participant pool includes college students from diverse fields such as Computer Science, Library and Information Science, and Public Health. Additionally, we incorporated crowd workers to ensure an even broader user spectrum. However, we set a qualification with an age range of 18-25 for crowd workers to facilitate a comparative analysis between the two groups. 

\subsection{Analysis}
For this small-scale user study, we utilized a descriptive analysis by presenting the tasks users performed on our platform and explaining how the platform influenced user experience and assisted them in task completion. In addition, we delved into the interview data as case studies to illuminate users' experiences and insights.

\section{Results}
\subsection{Participants and tasks}
As a result, 16 participants enrolled in the user study, comprising 8 college students and 8 crowd workers. The college students came from various academic backgrounds, including computer science, library/information science, and public health, ranging from sophomores to graduates. The crowd workers specialized in fields such as information technology and business, and were either pursuing or had already obtained their bachelor’s degrees. Out of the 16 participants, six completed the tasks according to their task plans and participated in the interview (3 college students and 3 crowd workers), while the remainder finished at least the warm-up task.

Table \ref{tab:exp} presents the average results of the tasks. Excluding the warm-up task, there were 29 tasks in total, comprising 10 short tasks, 13 medium tasks, and 6 long tasks. There were notable differences between the college student group and the crowd worker group, especially concerning the numbers of prompts and used prompt suggestions, and task duration. College students submitted approximately 5 to 6 prompts in short or medium tasks, though the duration for medium tasks was about double that of short tasks. They submitted over 20 prompts in the long task, completing the task in almost two days. In all task lengths, they adopted about one prompt from the suggestions in average. Conversely, crowd workers spent more time and prompts on short tasks than college students did, but they spent less time on medium and long tasks. This discrepancy could stem from their reliance on prompt suggestions, as nearly all their prompts were derived from these suggestions. Regarding the conversation ratings, college students had a positive experience (high usefulness, explanation utility, and satisfaction) in short and medium tasks but a moderate experience in long tasks. Crowd workers generally had a positive experience, except for the moderate explanation utility in long tasks.

\begin{table}
\caption{Average results of the tasks and user experience.}
\label{tab:exp}
\centering
\footnotesize
\begin{tabular}{l|ccc|ccc} \hline
Participant group & \multicolumn{3}{c|}{College student} & \multicolumn{3}{c}{Crowd worker} \\ \hline
Expected length & Short & Median & Long & Short & Median & Long \\ 
Task count & 7 & 6 & 3 & 3 & 7 & 3 \\
\hline
\# Prompt & 5.9 & 5.2 & 22.7 & 12.5 & 12.5 & 25.3 \\
\# Suggestion & 1.2 & 0.8 & 1 & 12.5 & 12 & 23 \\
Duration (min)* & 223.5 & 445.2 & 2488.6 & 558.3 & 31.5 & 32.9 \\
Usefulness & 3.8 & 4.2 & 3.2 & 4.3 & 4.2 & 4 \\
Explanation utility & 4 & 4.1 & 3.2 & 4.2 & 3.8 & 2.8 \\
Satisfaction & 4 & 4.2 & 3.3 & 4.3 & 4.5 & 4.2 \\ \hline
\end{tabular}
\begin{minipage}{\linewidth}
\vspace{0.1cm}
\scriptsize * This duration reflects the gap between the start and end of the task, not the exact amount of time spent on the task.
\end{minipage}
\end{table}

To gain deeper insight into the participants' experiences, table \ref{tab:task} provides a summary and examples of tasks topics and types, grouped by users and expected task length. College students (especially computer science students) engaged in tasks that included learning new topics (in short and medium tasks) and solving specific programming problems (in medium and long tasks). In contrast, crowd workers did not specify clear task topics in the pre-task questionnaire. They used vague terms like "JK," "learning," and "developing," primarily intending to engage in casual conversations with ChatGPT. Consequently, based on their input and heavy reliance on the prompt suggestion function, those suggestions led the conversations towards topics such as "J.K. Rowling", "online learning platforms", and "learning programming languages".

\begin{table*}
\caption{Summary and examples of task topics and types grouped by expected length and participant group.}
\label{tab:task}
\footnotesize
\centering
\begin{tabular}{l|l|l|l} \hline
Expected length & \multicolumn{1}{c|}{Short} & \multicolumn{1}{c|}{Median} & \multicolumn{1}{c}{Long} \\ \hline
Participant group & \multicolumn{3}{c}{College student  } \\ \hline
Topic & \begin{tabular}[c]{@{}l@{}}Learning a language, \\Grammar Checking, \\Basic programming\end{tabular} & Specific programming problems & Specific programming problems \\ \hline
Type & \begin{tabular}[c]{@{}l@{}}Learning a new topic,\\Writing or generating text\end{tabular} & \begin{tabular}[c]{@{}l@{}}Learning a new topic,\\Developing or programming\end{tabular} & Developing or programming \\ \hline
Participant group & \multicolumn{3}{c}{Crowd worker} \\ \hline
Topic & (writing techniques) & \begin{tabular}[c]{@{}l@{}}JK (Rowling),\\Company market\end{tabular} & \begin{tabular}[c]{@{}l@{}}Learning (online learning platform),\\Developing (learning programming languages)\end{tabular} \\ \hline
Type & \begin{tabular}[c]{@{}l@{}}Casual conversation,\\(Learning anew topic)\end{tabular} & \begin{tabular}[c]{@{}l@{}}Casual conversation,\\(Learning a new topic)\end{tabular} & \begin{tabular}[c]{@{}l@{}}Casual conversation, \\Learning a new topic\end{tabular} \\ \hline
\end{tabular}
\begin{minipage}{\linewidth}
\vspace{0.1cm}
\scriptsize 
Task topics and types in "()" provide clarifications for ambiguous topics or types that users entered in the pre-task questionnaire, as further inferred from actual conversations.
\end{minipage}
\end{table*}

\subsection{Interview case study on task experience}
We further present insights from the interview as case studies, examining the impact of supportive functions on user experience and task completion. Table 3 outlines participants' backgrounds, prior experiences with ChatGPT, task experiences during this study, and insights. We interviewed three computer science college students, P1, P2, and P3, all of whom showed considerable enthusiasm and engagement in both the tasks and subsequent interviews. Additionally, we interviewed crowd workers P4, P5, and P6. We delve into detailed feedback from P1, P2, and P3 and summarize the concerns raised by P4, P5, and P6.
\begin{table*}
\caption{Summary of user experience and insights from the interview.}
\label{tab:interview}
\centering
\footnotesize
\begin{tabular}{>{\hspace{0pt}}m{0.07\linewidth}|>{\hspace{0pt}}m{0.2\linewidth}|>{\hspace{0pt}}m{0.25\linewidth}|>{\hspace{0pt}}m{0.2\linewidth}|>{\hspace{0pt}}m{0.16\linewidth}} \hline
Participant & \multicolumn{1}{>{\centering\hspace{0pt}}m{0.2\linewidth}|}{P1} & \multicolumn{1}{>{\centering\hspace{0pt}}m{0.25\linewidth}|}{P2} & \multicolumn{1}{>{\centering\hspace{0pt}}m{0.2\linewidth}|}{P3} & \multicolumn{1}{>{\centering\arraybackslash\hspace{0pt}}m{0.16\linewidth}}{P4 P5 P6} \\ \hline
Background & Computer science senior & Computer science sophomore & Computer science sophomore & Crowd workers in information technology fields \\ \hline
Previous \par{}experience & ChatGPT not compatible with human experts & ChatGPT with unexpected performance & Miscommunication between the user and ChatGPT & Using ChatGPT for text generation work \\ \hline
Task in this study & Generating Haikus,  \par{}Internship specific question, \par{}Specific Coding problem & Learning a game dev tool,  \par{}Using a game dev package, \par{}Specific coding problem & Resume revision & Casual conversation, \par{}learning a topic \\ \hline
Experience in this study & • Detailed pre-task perceptions for \textbf{guiding expectations}; \par{}• Prompt explanations for \textbf{balancing expectations}; \par{}• Prompt suggestions for curiosity or unfamiliarity; \par{}• Task-level conversation & • Overestimated familiarity recognition through \textbf{retrospection}; \par{}• Prompt suggestions for initializing the conversation; \par{}• Prompt suggestion not applicable for complex tasks; \par{}• Explanations for \textbf{prompt refining}; \par{}• Final prompt development or \textbf{final solution} guidance & • Paraphrasing issues in explanations; \par{}• Distracting issues in prompt suggestions; \par{}• Prompt suggestions for \textbf{new and diverse} information; \par{}• Efforts in refining desired output & • Prompt suggestion reliance; \par{}• Satisfaction \\ \hline
Insight & • Expectation management\par{}• Task reliability\par{}• Prompt refining willingness\par{}• Diverse prompt suggestions\par{}• Task-focused interface & • Dynamic perceptions\par{}• Expectation management\par{}• Task reliability\par{}• Explainability for prompt refining\par{}• Interactive prompt refining process & • Diverse explanation,\par{}• Diverse prompt suggestions\par{}• Interactive prompt refining process & • Misuse of prompt suggestions \\ \hline
\end{tabular}
\end{table*}

P1, a self-identified expert of ChatGPT, considering ChatGPT as useful but not compatible with human experts. In this study, P1 recognized the value of perception articulation in "guiding expectations". In addition, the conversation explanations helped P1 "balance expectations" and satisfaction. For instance, P1 previously experienced ChatGPT's shortcomings in understand the specific rules of haikus, and they (a gender-neutral alternative to he/she) had a low expectation and started with some simple rules of haikus in this study. Surprisingly, P1 found the generated haikus followed the correct rules. However, inconsistencies arose when P1 increased the rule complexity. After reviewing the conversation explanations, P1 adjusted their expectations, acknowledging ChatGPT's limitations but remaining satisfied. P1 also noticed that explanations were more detailed for tasks with low familiarity but more concise and “straight to the point” for more familiar topics. However, P1 found the prompt suggestion function sometimes misaligned with their intended task direction, particularly in specific programming problems. Nonetheless, P1 appreciated the guidance from suggested prompts when exploring unfamiliar content, especially in the internship related question. They also highlighted the study platform's effectiveness in maintaining focus on specific tasks, thereby enhancing engagement and efficiency.

P2 previously found ChatGPT unexpectedly efficient in suggesting coding solutions. However, they did not anticipate fully completing tasks from ChatGPT's responses but expected to "get a broad idea" to guide their own solution development. In this study, P2 recognized how the platform interpreted task context and user perceptions in the pre-task questionnaire. For example, when encountering challenges that ChatGPT’s response exceeded P2's knowledge, they realized overestimated familiarity of a topic and reassessed it. The conversation explanation function allowed P2 to understand where any miscommunication began and to refine prompts. For example, in the complex C++ coding task, P2 gained confidence and refined prompts through iterative interactions and could use ChatGPT to "build all remembered functions in one prompt", attributing this improvement to learning from the explanations. However, P2 became dissatisfied when the code generated by ChatGPT was incompatible with an updated game development package. They then realized that ChatGPT had limited knowledge on this development package and discontinued the conversation for alternative resources. Regarding the prompt suggestions, P2 found them useful in initializing short tasks, but overly broad and generalized in the longer or more complex tasks.

P3 self-identified as proficient with ChatGPT, but they experienced miscommunication with ChatGPT’s interpretations. In this study, P3 provided unique insights when revising their resume. P3 observed redundancy and paraphrasing issues among the explanation options. For the prompt suggestions, P3 felt that those suggestions could sometimes distract from their initial intents. Nevertheless, P3 acknowledged the suggestions' potential to unveil new perspectives, such as unconsidered resume layouts. P3 also emphasized the effort required to modify ChatGPT’s output to their preferred style and content.

Regarding the crowd workers (P4, P5, and P6) involved in IT-related fields, their previous interaction with ChatGPT was primarily work-related. In this study, as per the tasks summarized in Tables 1 and 2, they appeared hurried in their completion, heavily relying on prompt suggestions, possibly due to their focus on Human Intelligence Tasks (HITs). Their feedback during interviews remained nonspecific, centered more around task completion for compensation rather than the study’s investigative purpose. It seems they misunderstood the task purpose of this study, or they were outliers of target users, which raised concerns discussed in the next section. 

\section{Discussion and conclusion}
In closing discussion, we incorporate insights derived from our study, focusing on enhancing human-LLM interaction and identifying prevalent challenges during these interactions.

\textit{Insights and Implications for System Evaluation}: The interview insights suggest distinct evaluation metrics at both the system and user ends. For the system, we can propose task reliability, which reflects the LLM's deep understanding and capability of task-specific knowledge beyond the text completion. This reliability could be evaluated through the LLM's ability to describe different levels of task familiarity and complexity for better recognizing the user's task states given the task familiarity and complexity levels. Another evaluation aspect could be conversation explainability, reflecting the LLM's ability to interpret the conversation from a task-aware perspective. Enhancements of both task reliability and explainability might include graph-based methods to represent the task in a knowledge structure with contextual information that enhances situational awareness \cite{sarkar2023representing, wang2019explainable}. On the user end, metrics on task engagement could be measured by users' willingness to refine prompts and be influenced by the expectation of LLM's capability \cite{zamfirescu2023johnny, liu2018satisfaction}.

\textit{Proactive User Interface and System Design}: Our study offers insights for proactive user interface and system design, which extends previous work on proactive IR \cite{liu2019proactive}. Our approach involved utilizing questionnaires and specific interface tabs, yet the need for more seamless integration emerged. Following established AI interaction guidelines, a separate interface, such as a sidebar or secondary tab, could be deployed to facilitate supportive functions without interruptions on the ongoing conversation with ChatGPT. This proactive interface could also adopt a conversation-based mechanism for capturing user perceptions and provide suggestions, assisting the main ChatGPT interaction. This system holds the potential to leverage a fine-tuned LLM, for analyzing ongoing conversations and offering task-aware recommendations \cite{liao2023proactivea, liao2023proactiveb, shah2023taking, liu2019task}. This insight represents a step towards more intuitive, assistive AI, catering to diverse user needs without over-complicating the interaction.

\textit{Recognizing the Unique Position of Crowd Workers}: This study revealed unexpected findings concerning crowd workers' interactions with ChatGPT. These anomalies may not solely be attributed to users' misunderstanding of the study's purpose but also possibly to the inherent challenges these users faced in formulating their own tasks. HITs for crowd workers are usually straightforward and repetitive tasks, like data labelling, even though they still require instructions and a gold standard to ensure accurate labeling \cite{thomas2023large}. Although they can produce text resembling task plans to meet the HIT criteria, this text does not necessarily mirror their real information needs \cite{ross2010crowdworkers, steiner2021crowdsourcing}. Formulating their "own" tasks or even identifying their information needs may be internally complex tasks. Additionally, the emphasis on task-centric LLM in this study might overshadow challenges with exploratory or open-ended user intentions. Therefore, it is crucial to comprehend the diverse information needs and usage contexts among more diverse user categories, so that the proactive system should be intuitively informative and inclusive, rather than restrictively functional for certain user groups \cite{shah2023envisioning}.

\textit{Limitations and Future Directions}: Acknowledging our study's constraints in its limited scale, we advocate for expansive user studies with participants from diverse communities, populations, and backgrounds.
Beginning for the prototype system in this study, future work could involve improving the interface and system design, conducting prompt template ablation studies, and exploring LLM fine-turning for developing task support affordances. Further discussions might explore task automation and copilot, with the challenge of balancing user engagement in information seeking and learning, while providing efficient, single-prompt task resolution.

In conclusion, our user study delved into supportive functions for ChatGPT, bolstering user experiences and task completion. The results indicated that these functions effectively guided users in managing expectations, reducing cognitive load, better refining prompts, and increasing user engagement. This investigation markedly advances our understanding of how to leverage LLMs for proactive IS systems. Moreover, it sheds light on the evaluation of human-LLM interactions and highlights the potential oversight of certain user groups' unique challenges in the era of Generative AI.

\section{Acknowledgment}
This work is supported by the National Science Foundation (NSF) Award IIS-2106152, a grant from the Seed Funding Program of the Data Institute for Societal Challenges, the University of Oklahoma, and a fund from Microsoft for Startups Founders Hub. Any opinions, findings, conclusions or recommendations expressed in this material are those of the authors and do not necessarily reflect those of the sponsors.

%%
%% The next two lines define the bibliography style to be used, and
%% the bibliography file.
% \balance
\bibliographystyle{ACM-Reference-Format}
\bibliography{sample-base}

%%% -*-BibTeX-*-
%%% Do NOT edit. File created by BibTeX with style
%%% ACM-Reference-Format-Journals [18-Jan-2012].

\begin{thebibliography}{35}

%%% ====================================================================
%%% NOTE TO THE USER: you can override these defaults by providing
%%% customized versions of any of these macros before the \bibliography
%%% command.  Each of them MUST provide its own final punctuation,
%%% except for \shownote{}, \showDOI{}, and \showURL{}.  The latter two
%%% do not use final punctuation, in order to avoid confusing it with
%%% the Web address.
%%%
%%% To suppress output of a particular field, define its macro to expand
%%% to an empty string, or better, \unskip, like this:
%%%
%%% \newcommand{\showDOI}[1]{\unskip}   % LaTeX syntax
%%%
%%% \def \showDOI #1{\unskip}           % plain TeX syntax
%%%
%%% ====================================================================

\ifx \showCODEN    \undefined \def \showCODEN     #1{\unskip}     \fi
\ifx \showDOI      \undefined \def \showDOI       #1{#1}\fi
\ifx \showISBNx    \undefined \def \showISBNx     #1{\unskip}     \fi
\ifx \showISBNxiii \undefined \def \showISBNxiii  #1{\unskip}     \fi
\ifx \showISSN     \undefined \def \showISSN      #1{\unskip}     \fi
\ifx \showLCCN     \undefined \def \showLCCN      #1{\unskip}     \fi
\ifx \shownote     \undefined \def \shownote      #1{#1}          \fi
\ifx \showarticletitle \undefined \def \showarticletitle #1{#1}   \fi
\ifx \showURL      \undefined \def \showURL       {\relax}        \fi
% The following commands are used for tagged output and should be
% invisible to TeX
\providecommand\bibfield[2]{#2}
\providecommand\bibinfo[2]{#2}
\providecommand\natexlab[1]{#1}
\providecommand\showeprint[2][]{arXiv:#2}

\bibitem[Amershi et~al\mbox{.}(2019)]%
        {amershi2019guidelines}
\bibfield{author}{\bibinfo{person}{Saleema Amershi}, \bibinfo{person}{Dan Weld}, \bibinfo{person}{Mihaela Vorvoreanu}, \bibinfo{person}{Adam Fourney}, \bibinfo{person}{Besmira Nushi}, \bibinfo{person}{Penny Collisson}, \bibinfo{person}{Jina Suh}, \bibinfo{person}{Shamsi Iqbal}, \bibinfo{person}{Paul~N Bennett}, \bibinfo{person}{Kori Inkpen}, {et~al\mbox{.}}} \bibinfo{year}{2019}\natexlab{}.
\newblock \showarticletitle{Guidelines for human-AI interaction}. In \bibinfo{booktitle}{\emph{Proceedings of the 2019 chi conference on human factors in computing systems}}. \bibinfo{pages}{1--13}.
\newblock


\bibitem[Azzopardi(2021)]%
        {azzopardi2021cognitive}
\bibfield{author}{\bibinfo{person}{Leif Azzopardi}.} \bibinfo{year}{2021}\natexlab{}.
\newblock \showarticletitle{Cognitive biases in search: a review and reflection of cognitive biases in Information Retrieval}. In \bibinfo{booktitle}{\emph{Proceedings of the 2021 conference on human information interaction and retrieval}}. \bibinfo{pages}{27--37}.
\newblock


\bibitem[Bahrini et~al\mbox{.}(2023)]%
        {bahrini2023chatgpt}
\bibfield{author}{\bibinfo{person}{Aram Bahrini}, \bibinfo{person}{Mohammadsadra Khamoshifar}, \bibinfo{person}{Hossein Abbasimehr}, \bibinfo{person}{Robert~J Riggs}, \bibinfo{person}{Maryam Esmaeili}, \bibinfo{person}{Rastin~Mastali Majdabadkohne}, {and} \bibinfo{person}{Morteza Pasehvar}.} \bibinfo{year}{2023}\natexlab{}.
\newblock \showarticletitle{ChatGPT: Applications, opportunities, and threats}. In \bibinfo{booktitle}{\emph{2023 Systems and Information Engineering Design Symposium (SIEDS)}}. IEEE, \bibinfo{pages}{274--279}.
\newblock


\bibitem[Baidoo-Anu and Ansah(2023)]%
        {baidoo2023education}
\bibfield{author}{\bibinfo{person}{David Baidoo-Anu} {and} \bibinfo{person}{Leticia~Owusu Ansah}.} \bibinfo{year}{2023}\natexlab{}.
\newblock \showarticletitle{Education in the era of generative artificial intelligence (AI): Understanding the potential benefits of ChatGPT in promoting teaching and learning}.
\newblock \bibinfo{journal}{\emph{Journal of AI}} \bibinfo{volume}{7}, \bibinfo{number}{1} (\bibinfo{year}{2023}), \bibinfo{pages}{52--62}.
\newblock


\bibitem[Brown et~al\mbox{.}(2020)]%
        {brown2020language}
\bibfield{author}{\bibinfo{person}{Tom Brown}, \bibinfo{person}{Benjamin Mann}, \bibinfo{person}{Nick Ryder}, \bibinfo{person}{Melanie Subbiah}, \bibinfo{person}{Jared~D Kaplan}, \bibinfo{person}{Prafulla Dhariwal}, \bibinfo{person}{Arvind Neelakantan}, \bibinfo{person}{Pranav Shyam}, \bibinfo{person}{Girish Sastry}, \bibinfo{person}{Amanda Askell}, {et~al\mbox{.}}} \bibinfo{year}{2020}\natexlab{}.
\newblock \showarticletitle{Language models are few-shot learners}.
\newblock \bibinfo{journal}{\emph{Advances in neural information processing systems}}  \bibinfo{volume}{33} (\bibinfo{year}{2020}), \bibinfo{pages}{1877--1901}.
\newblock


\bibitem[Ford(1995)]%
        {ford1995information}
\bibfield{author}{\bibinfo{person}{Barbara~J Ford}.} \bibinfo{year}{1995}\natexlab{}.
\newblock \showarticletitle{Information literacy as a barrier}.
\newblock \bibinfo{journal}{\emph{IFLA journal}} \bibinfo{volume}{21}, \bibinfo{number}{2} (\bibinfo{year}{1995}), \bibinfo{pages}{99--101}.
\newblock


\bibitem[Hassan et~al\mbox{.}(2014)]%
        {hassan2014struggling}
\bibfield{author}{\bibinfo{person}{Ahmed Hassan}, \bibinfo{person}{Ryen~W White}, \bibinfo{person}{Susan~T Dumais}, {and} \bibinfo{person}{Yi-Min Wang}.} \bibinfo{year}{2014}\natexlab{}.
\newblock \showarticletitle{Struggling or exploring? Disambiguating long search sessions}. In \bibinfo{booktitle}{\emph{Proceedings of the 7th ACM international conference on Web search and data mining}}. \bibinfo{pages}{53--62}.
\newblock


\bibitem[Kasneci et~al\mbox{.}(2023)]%
        {kasneci2023chatgpt}
\bibfield{author}{\bibinfo{person}{Enkelejda Kasneci}, \bibinfo{person}{Kathrin Se{\ss}ler}, \bibinfo{person}{Stefan K{\"u}chemann}, \bibinfo{person}{Maria Bannert}, \bibinfo{person}{Daryna Dementieva}, \bibinfo{person}{Frank Fischer}, \bibinfo{person}{Urs Gasser}, \bibinfo{person}{Georg Groh}, \bibinfo{person}{Stephan G{\"u}nnemann}, \bibinfo{person}{Eyke H{\"u}llermeier}, {et~al\mbox{.}}} \bibinfo{year}{2023}\natexlab{}.
\newblock \showarticletitle{ChatGPT for good? On opportunities and challenges of large language models for education}.
\newblock \bibinfo{journal}{\emph{Learning and individual differences}}  \bibinfo{volume}{103} (\bibinfo{year}{2023}), \bibinfo{pages}{102274}.
\newblock


\bibitem[Kim et~al\mbox{.}(2023)]%
        {kim2023help}
\bibfield{author}{\bibinfo{person}{Sunnie~SY Kim}, \bibinfo{person}{Elizabeth~Anne Watkins}, \bibinfo{person}{Olga Russakovsky}, \bibinfo{person}{Ruth Fong}, {and} \bibinfo{person}{Andr{\'e}s Monroy-Hern{\'a}ndez}.} \bibinfo{year}{2023}\natexlab{}.
\newblock \showarticletitle{" Help Me Help the AI": Understanding How Explainability Can Support Human-AI Interaction}. In \bibinfo{booktitle}{\emph{Proceedings of the 2023 CHI Conference on Human Factors in Computing Systems}}. \bibinfo{pages}{1--17}.
\newblock


\bibitem[Lee et~al\mbox{.}(2022)]%
        {lee2022evaluating}
\bibfield{author}{\bibinfo{person}{Mina Lee}, \bibinfo{person}{Megha Srivastava}, \bibinfo{person}{Amelia Hardy}, \bibinfo{person}{John Thickstun}, \bibinfo{person}{Esin Durmus}, \bibinfo{person}{Ashwin Paranjape}, \bibinfo{person}{Ines Gerard-Ursin}, \bibinfo{person}{Xiang~Lisa Li}, \bibinfo{person}{Faisal Ladhak}, \bibinfo{person}{Frieda Rong}, {et~al\mbox{.}}} \bibinfo{year}{2022}\natexlab{}.
\newblock \showarticletitle{Evaluating human-language model interaction}.
\newblock \bibinfo{journal}{\emph{arXiv preprint arXiv:2212.09746}} (\bibinfo{year}{2022}).
\newblock


\bibitem[Liao et~al\mbox{.}(2023a)]%
        {liao2023proactivea}
\bibfield{author}{\bibinfo{person}{Lizi Liao}, \bibinfo{person}{Grace~Hui Yang}, {and} \bibinfo{person}{Chirag Shah}.} \bibinfo{year}{2023}\natexlab{a}.
\newblock \showarticletitle{Proactive conversational agents}. In \bibinfo{booktitle}{\emph{Proceedings of the Sixteenth ACM International Conference on Web Search and Data Mining}}. \bibinfo{pages}{1244--1247}.
\newblock


\bibitem[Liao et~al\mbox{.}(2023b)]%
        {liao2023proactiveb}
\bibfield{author}{\bibinfo{person}{Lizi Liao}, \bibinfo{person}{Grace~Hui Yang}, {and} \bibinfo{person}{Chirag Shah}.} \bibinfo{year}{2023}\natexlab{b}.
\newblock \showarticletitle{Proactive Conversational Agents in the Post-ChatGPT World}. In \bibinfo{booktitle}{\emph{Proceedings of the 46th International ACM SIGIR Conference on Research and Development in Information Retrieval}}. \bibinfo{pages}{3452--3455}.
\newblock


\bibitem[Liu et~al\mbox{.}(2019)]%
        {liu2019task}
\bibfield{author}{\bibinfo{person}{Jiqun Liu}, \bibinfo{person}{Matthew Mitsui}, \bibinfo{person}{Nicholas~J Belkin}, {and} \bibinfo{person}{Chirag Shah}.} \bibinfo{year}{2019}\natexlab{}.
\newblock \showarticletitle{Task, information seeking intentions, and user behavior: Toward a multi-level understanding of Web search}. In \bibinfo{booktitle}{\emph{Proceedings of the 2019 conference on human information interaction and retrieval}}. \bibinfo{pages}{123--132}.
\newblock


\bibitem[Liu et~al\mbox{.}(2020)]%
        {liu2020identifying}
\bibfield{author}{\bibinfo{person}{Jiqun Liu}, \bibinfo{person}{Shawon Sarkar}, {and} \bibinfo{person}{Chirag Shah}.} \bibinfo{year}{2020}\natexlab{}.
\newblock \showarticletitle{Identifying and predicting the states of complex search tasks}. In \bibinfo{booktitle}{\emph{Proceedings of the 2020 conference on human information interaction and retrieval}}. \bibinfo{pages}{193--202}.
\newblock


\bibitem[Liu and Shah(2019)]%
        {liu2019proactive}
\bibfield{author}{\bibinfo{person}{Jiqun Liu} {and} \bibinfo{person}{Chirag Shah}.} \bibinfo{year}{2019}\natexlab{}.
\newblock \showarticletitle{Proactive identification of query failure}.
\newblock \bibinfo{journal}{\emph{Proceedings of the Association for Information Science and Technology}} \bibinfo{volume}{56}, \bibinfo{number}{1} (\bibinfo{year}{2019}), \bibinfo{pages}{176--185}.
\newblock


\bibitem[Liu and Shah(2022)]%
        {liu2022leveraging}
\bibfield{author}{\bibinfo{person}{Jiqun Liu} {and} \bibinfo{person}{Chirag Shah}.} \bibinfo{year}{2022}\natexlab{}.
\newblock \showarticletitle{Leveraging user interaction signals and task state information in adaptively optimizing usefulness-oriented search sessions}. In \bibinfo{booktitle}{\emph{Proceedings of the 22nd ACM/IEEE joint conference on digital libraries}}. \bibinfo{pages}{1--11}.
\newblock


\bibitem[Liu et~al\mbox{.}(2018)]%
        {liu2018satisfaction}
\bibfield{author}{\bibinfo{person}{Mengyang Liu}, \bibinfo{person}{Yiqun Liu}, \bibinfo{person}{Jiaxin Mao}, \bibinfo{person}{Cheng Luo}, \bibinfo{person}{Min Zhang}, {and} \bibinfo{person}{Shaoping Ma}.} \bibinfo{year}{2018}\natexlab{}.
\newblock \showarticletitle{" Satisfaction with Failure" or" Unsatisfied Success" Investigating the Relationship between Search Success and User Satisfaction}. In \bibinfo{booktitle}{\emph{Proceedings of the 2018 world wide web conference}}. \bibinfo{pages}{1533--1542}.
\newblock


\bibitem[Liu et~al\mbox{.}(2023)]%
        {liu2023pre}
\bibfield{author}{\bibinfo{person}{Pengfei Liu}, \bibinfo{person}{Weizhe Yuan}, \bibinfo{person}{Jinlan Fu}, \bibinfo{person}{Zhengbao Jiang}, \bibinfo{person}{Hiroaki Hayashi}, {and} \bibinfo{person}{Graham Neubig}.} \bibinfo{year}{2023}\natexlab{}.
\newblock \showarticletitle{Pre-train, prompt, and predict: A systematic survey of prompting methods in natural language processing}.
\newblock \bibinfo{journal}{\emph{Comput. Surveys}} \bibinfo{volume}{55}, \bibinfo{number}{9} (\bibinfo{year}{2023}), \bibinfo{pages}{1--35}.
\newblock


\bibitem[Odijk et~al\mbox{.}(2015)]%
        {odijk2015struggling}
\bibfield{author}{\bibinfo{person}{Daan Odijk}, \bibinfo{person}{Ryen~W White}, \bibinfo{person}{Ahmed Hassan~Awadallah}, {and} \bibinfo{person}{Susan~T Dumais}.} \bibinfo{year}{2015}\natexlab{}.
\newblock \showarticletitle{Struggling and success in web search}. In \bibinfo{booktitle}{\emph{Proceedings of the 24th ACM International on Conference on Information and Knowledge Management}}. \bibinfo{pages}{1551--1560}.
\newblock


\bibitem[Ouyang et~al\mbox{.}(2022)]%
        {ouyang2022training}
\bibfield{author}{\bibinfo{person}{Long Ouyang}, \bibinfo{person}{Jeffrey Wu}, \bibinfo{person}{Xu Jiang}, \bibinfo{person}{Diogo Almeida}, \bibinfo{person}{Carroll Wainwright}, \bibinfo{person}{Pamela Mishkin}, \bibinfo{person}{Chong Zhang}, \bibinfo{person}{Sandhini Agarwal}, \bibinfo{person}{Katarina Slama}, \bibinfo{person}{Alex Ray}, {et~al\mbox{.}}} \bibinfo{year}{2022}\natexlab{}.
\newblock \showarticletitle{Training language models to follow instructions with human feedback}.
\newblock \bibinfo{journal}{\emph{Advances in Neural Information Processing Systems}}  \bibinfo{volume}{35} (\bibinfo{year}{2022}), \bibinfo{pages}{27730--27744}.
\newblock


\bibitem[Ross et~al\mbox{.}(2010)]%
        {ross2010crowdworkers}
\bibfield{author}{\bibinfo{person}{Joel Ross}, \bibinfo{person}{Lilly Irani}, \bibinfo{person}{M~Six Silberman}, \bibinfo{person}{Andrew Zaldivar}, {and} \bibinfo{person}{Bill Tomlinson}.} \bibinfo{year}{2010}\natexlab{}.
\newblock \showarticletitle{Who are the crowdworkers? Shifting demographics in Mechanical Turk}.
\newblock In \bibinfo{booktitle}{\emph{CHI'10 extended abstracts on Human factors in computing systems}}. \bibinfo{pages}{2863--2872}.
\newblock


\bibitem[Sarkar et~al\mbox{.}(2023)]%
        {sarkar2023representing}
\bibfield{author}{\bibinfo{person}{Shawon Sarkar}, \bibinfo{person}{Maryam Amirizaniani}, {and} \bibinfo{person}{Chirag Shah}.} \bibinfo{year}{2023}\natexlab{}.
\newblock \showarticletitle{Representing Tasks with a Graph-Based Method for Supporting Users in Complex Search Tasks}. In \bibinfo{booktitle}{\emph{Proceedings of the 2023 Conference on Human Information Interaction and Retrieval}}. \bibinfo{pages}{378--382}.
\newblock


\bibitem[Savolainen(2015)]%
        {savolainen2015cognitive}
\bibfield{author}{\bibinfo{person}{Reijo Savolainen}.} \bibinfo{year}{2015}\natexlab{}.
\newblock \showarticletitle{Cognitive barriers to information seeking: A conceptual analysis}.
\newblock \bibinfo{journal}{\emph{Journal of Information Science}} \bibinfo{volume}{41}, \bibinfo{number}{5} (\bibinfo{year}{2015}), \bibinfo{pages}{613--623}.
\newblock


\bibitem[SHAH and BENDER(2023)]%
        {shah2023envisioning}
\bibfield{author}{\bibinfo{person}{CHIRAG SHAH} {and} \bibinfo{person}{EMILY~M BENDER}.} \bibinfo{year}{2023}\natexlab{}.
\newblock \showarticletitle{Envisioning Information Access Systems: What Makes for Good Tools and a Healthy Web?}
\newblock  (\bibinfo{year}{2023}).
\newblock


\bibitem[Shah et~al\mbox{.}(2023)]%
        {shah2023taking}
\bibfield{author}{\bibinfo{person}{Chirag Shah}, \bibinfo{person}{Ryen White}, \bibinfo{person}{Paul Thomas}, \bibinfo{person}{Bhaskar Mitra}, \bibinfo{person}{Shawon Sarkar}, {and} \bibinfo{person}{Nicholas Belkin}.} \bibinfo{year}{2023}\natexlab{}.
\newblock \showarticletitle{Taking search to task}. In \bibinfo{booktitle}{\emph{Proceedings of the 2023 Conference on Human Information Interaction and Retrieval}}. \bibinfo{pages}{1--13}.
\newblock


\bibitem[Skjuve et~al\mbox{.}(2023)]%
        {skjuve2023user}
\bibfield{author}{\bibinfo{person}{Marita Skjuve}, \bibinfo{person}{Asbj{\o}rn F{\o}lstad}, {and} \bibinfo{person}{Petter~Bae Brandtzaeg}.} \bibinfo{year}{2023}\natexlab{}.
\newblock \showarticletitle{The user experience of ChatGPT: Findings from a questionnaire study of early users}. In \bibinfo{booktitle}{\emph{Proceedings of the 5th International Conference on Conversational User Interfaces}}. \bibinfo{pages}{1--10}.
\newblock


\bibitem[Steiner et~al\mbox{.}(2021)]%
        {steiner2021crowdsourcing}
\bibfield{author}{\bibinfo{person}{Manuel Steiner}, \bibinfo{person}{Damiano Spina}, \bibinfo{person}{Falk Scholer}, {and} \bibinfo{person}{Lawrence Cavedon}.} \bibinfo{year}{2021}\natexlab{}.
\newblock \showarticletitle{Crowdsourcing Backstories for Complex Task-Based Search}. In \bibinfo{booktitle}{\emph{Proceedings of the 25th Australasian Document Computing Symposium}}. \bibinfo{pages}{1--6}.
\newblock


\bibitem[Thomas et~al\mbox{.}(2023)]%
        {thomas2023large}
\bibfield{author}{\bibinfo{person}{Paul Thomas}, \bibinfo{person}{Seth Spielman}, \bibinfo{person}{Nick Craswell}, {and} \bibinfo{person}{Bhaskar Mitra}.} \bibinfo{year}{2023}\natexlab{}.
\newblock \showarticletitle{Large language models can accurately predict searcher preferences}.
\newblock \bibinfo{journal}{\emph{arXiv preprint arXiv:2309.10621}} (\bibinfo{year}{2023}).
\newblock


\bibitem[Tlili et~al\mbox{.}(2023)]%
        {tlili2023if}
\bibfield{author}{\bibinfo{person}{Ahmed Tlili}, \bibinfo{person}{Boulus Shehata}, \bibinfo{person}{Michael~Agyemang Adarkwah}, \bibinfo{person}{Aras Bozkurt}, \bibinfo{person}{Daniel~T Hickey}, \bibinfo{person}{Ronghuai Huang}, {and} \bibinfo{person}{Brighter Agyemang}.} \bibinfo{year}{2023}\natexlab{}.
\newblock \showarticletitle{What if the devil is my guardian angel: ChatGPT as a case study of using chatbots in education}.
\newblock \bibinfo{journal}{\emph{Smart Learning Environments}} \bibinfo{volume}{10}, \bibinfo{number}{1} (\bibinfo{year}{2023}), \bibinfo{pages}{15}.
\newblock


\bibitem[Wang and Liu(2022)]%
        {wang2022investigating}
\bibfield{author}{\bibinfo{person}{Ben Wang} {and} \bibinfo{person}{Jiqun Liu}.} \bibinfo{year}{2022}\natexlab{}.
\newblock \showarticletitle{Investigating the Relationship between In-Situ User Expectations and Web Search Behavior}.
\newblock \bibinfo{journal}{\emph{Proceedings of the Association for Information Science and Technology}} \bibinfo{volume}{59}, \bibinfo{number}{1} (\bibinfo{year}{2022}), \bibinfo{pages}{827--829}.
\newblock


\bibitem[Wang and Liu(2023)]%
        {wang2023investigating}
\bibfield{author}{\bibinfo{person}{Ben Wang} {and} \bibinfo{person}{Jiqun Liu}.} \bibinfo{year}{2023}\natexlab{}.
\newblock \showarticletitle{Investigating the role of in-situ user expectations in Web search}.
\newblock \bibinfo{journal}{\emph{Information Processing \& Management}} \bibinfo{volume}{60}, \bibinfo{number}{3} (\bibinfo{year}{2023}), \bibinfo{pages}{103300}.
\newblock


\bibitem[Wang et~al\mbox{.}(2019)]%
        {wang2019explainable}
\bibfield{author}{\bibinfo{person}{Xiang Wang}, \bibinfo{person}{Dingxian Wang}, \bibinfo{person}{Canran Xu}, \bibinfo{person}{Xiangnan He}, \bibinfo{person}{Yixin Cao}, {and} \bibinfo{person}{Tat-Seng Chua}.} \bibinfo{year}{2019}\natexlab{}.
\newblock \showarticletitle{Explainable reasoning over knowledge graphs for recommendation}. In \bibinfo{booktitle}{\emph{Proceedings of the AAAI conference on artificial intelligence}}, Vol.~\bibinfo{volume}{33}. \bibinfo{pages}{5329--5336}.
\newblock


\bibitem[Xu et~al\mbox{.}(2019)]%
        {xu2019explainable}
\bibfield{author}{\bibinfo{person}{Feiyu Xu}, \bibinfo{person}{Hans Uszkoreit}, \bibinfo{person}{Yangzhou Du}, \bibinfo{person}{Wei Fan}, \bibinfo{person}{Dongyan Zhao}, {and} \bibinfo{person}{Jun Zhu}.} \bibinfo{year}{2019}\natexlab{}.
\newblock \showarticletitle{Explainable AI: A brief survey on history, research areas, approaches and challenges}. In \bibinfo{booktitle}{\emph{Natural Language Processing and Chinese Computing: 8th CCF International Conference, NLPCC 2019, Dunhuang, China, October 9--14, 2019, Proceedings, Part II 8}}. Springer, \bibinfo{pages}{563--574}.
\newblock


\bibitem[Zamfirescu-Pereira et~al\mbox{.}(2023)]%
        {zamfirescu2023johnny}
\bibfield{author}{\bibinfo{person}{JD Zamfirescu-Pereira}, \bibinfo{person}{Richmond~Y Wong}, \bibinfo{person}{Bjoern Hartmann}, {and} \bibinfo{person}{Qian Yang}.} \bibinfo{year}{2023}\natexlab{}.
\newblock \showarticletitle{Why Johnny can’t prompt: how non-AI experts try (and fail) to design LLM prompts}. In \bibinfo{booktitle}{\emph{Proceedings of the 2023 CHI Conference on Human Factors in Computing Systems}}. \bibinfo{pages}{1--21}.
\newblock


\bibitem[Zhai(2022)]%
        {zhai2022chatgpt}
\bibfield{author}{\bibinfo{person}{Xiaoming Zhai}.} \bibinfo{year}{2022}\natexlab{}.
\newblock \showarticletitle{ChatGPT user experience: Implications for education}.
\newblock \bibinfo{journal}{\emph{Available at SSRN 4312418}} (\bibinfo{year}{2022}).
\newblock


\end{thebibliography}

\end{document}